\documentclass[fleqn,12pt,twoside]{article}
\usepackage{espcrc1}

\usepackage{graphicx}
\usepackage[figuresright]{rotating}

\newcommand{\AmS}{{\protect\the\textfont2
  A\kern-.1667em\lower.5ex\hbox{M}\kern-.125emS}}

\title{Neutrino mean free path and in-medium nuclear interaction}

\author{J. Margueron%
\address{Istituto Nazionale di Fisica Nucleare, Sezione di Pisa, 56100 Pisa,
  Italy}
\address{GANIL CEA/DSM - CNRS/IN2P3 BP 5027 F-14076 Caen Cedex 5, France},
J. Navarro\address{IFIC (CSIC - Universidad de Valencia) Apdo. 22085,
E-46.071-Valencia, Spain}
and
N. Van Giai\address{Institut de Physique Nucl\'eaire, IN2P3-CNRS, F-91406
  Orsay Cedex, France}
}

\begin{document}

\maketitle

\begin{abstract}
Neutrinos produced during the collapse of a massive star are trapped in a
nuclear medium (the proto-neutron star). 
Typically, 
neutrino energies (10-100 MeV) are of the order
of nuclear giant resonances energies. Hence, neutrino propagation is
modified by the possibility of coherent scattering on nucleons.
We have compared the predictions of different nuclear interaction models.
It turns out that their main discrepancies are related to the density 
dependence of the k-effective mass as well as to the onset of 
instabilities as density increases.
This last point had led us to a systematic study of instabilities of 
infinite matter with effective Skyrme-type interactions. 
We have shown that for such interactions there is always a critical
density, above which the system becomes unstable. 
\end{abstract}

\vspace{0.5cm}
A proto-neutron star is formed during the collapse of a massive star. 
It is mainly composed of electrons, protons and neutrons but its core can
contain other particles like hyperons \cite{vida02}.
Neutrinos play a crucial role during the
energy liberation phase that follows a supernova collapse \cite{bur86}, as
the energy is carried away by neutrino scattering in the newly born
neutron star. During the first few seconds of the proto-neutron star, the matter is
neutrino rich and the leptonic number is close to that of nuclei. After
some tens seconds, at the end of the proto-neutron star, matter is neutrino poor
and the neutrino chemical potential is zero. 
It is important to have an accurate estimate of the mean free path of 
neutrinos in a nuclear medium where the density is equal or greater to the 
saturation nuclear matter density, the temperature a few tens of MeV and 
the proton fraction can reach 30\% of the baryonic density~\cite{red98}.
Here, we focus on the properties of nuclear matter for densities between 1 and
3 times $\rho_0$, the saturation value of symmetric nuclear matter. 
For such densities we can reasonably assume protons and neutrons as the 
only hadronic components.   
Our calculation of the neutrino mean free path is based on equations of state 
for pure neutron matter and asymmetric nuclear matter, employing   
effective Skyrme and Gogny nuclear interactions. 

\smallskip
We have considered pure neutron matter and asymmetric nuclear matter 
in $\beta$ equilibrium and we have calculated cross sections of reactions 
contributing to the neutrino mean free path.
Neutrino scattering off neutrons ($\nu + n \rightarrow \nu' + n'$) 
is the only contribution in pure neutron matter~\cite{iwa82,nav99}. 
In asymmetric matter, neutrinos can also
scatter off protons and electrons or be absorbed by neutrons 
($\nu + n \rightarrow e^- + p$). 
We have calculated these processes within the mean field approximation, 
and also in the framework of the 
Random Phase Approximation (RPA)~\cite{mar01b},
to consider the effects of nuclear 
correlations on the neutrino scattering.
\begin{figure}[ht]
\centering
\includegraphics[scale=0.25]{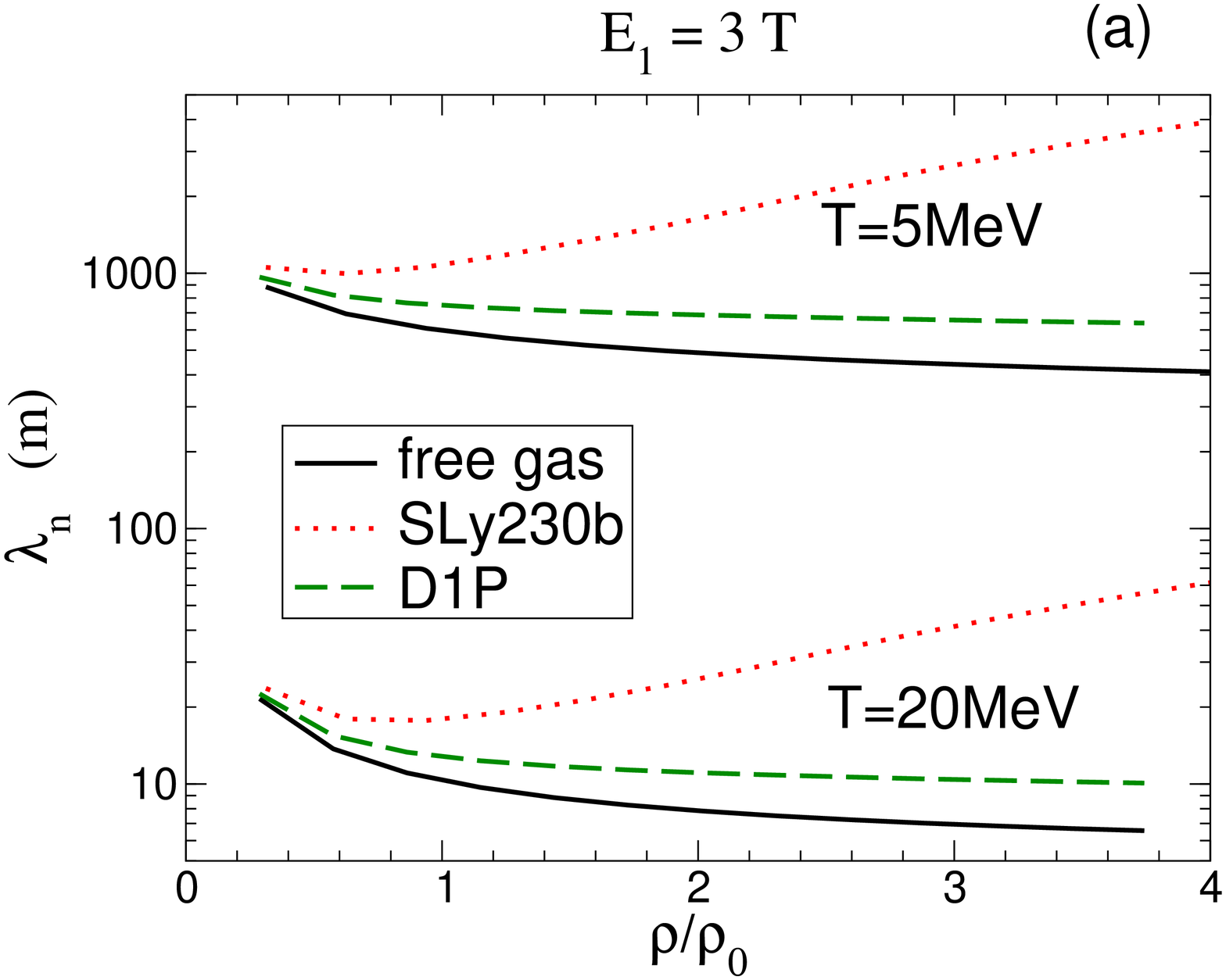}
\includegraphics[scale=0.25]{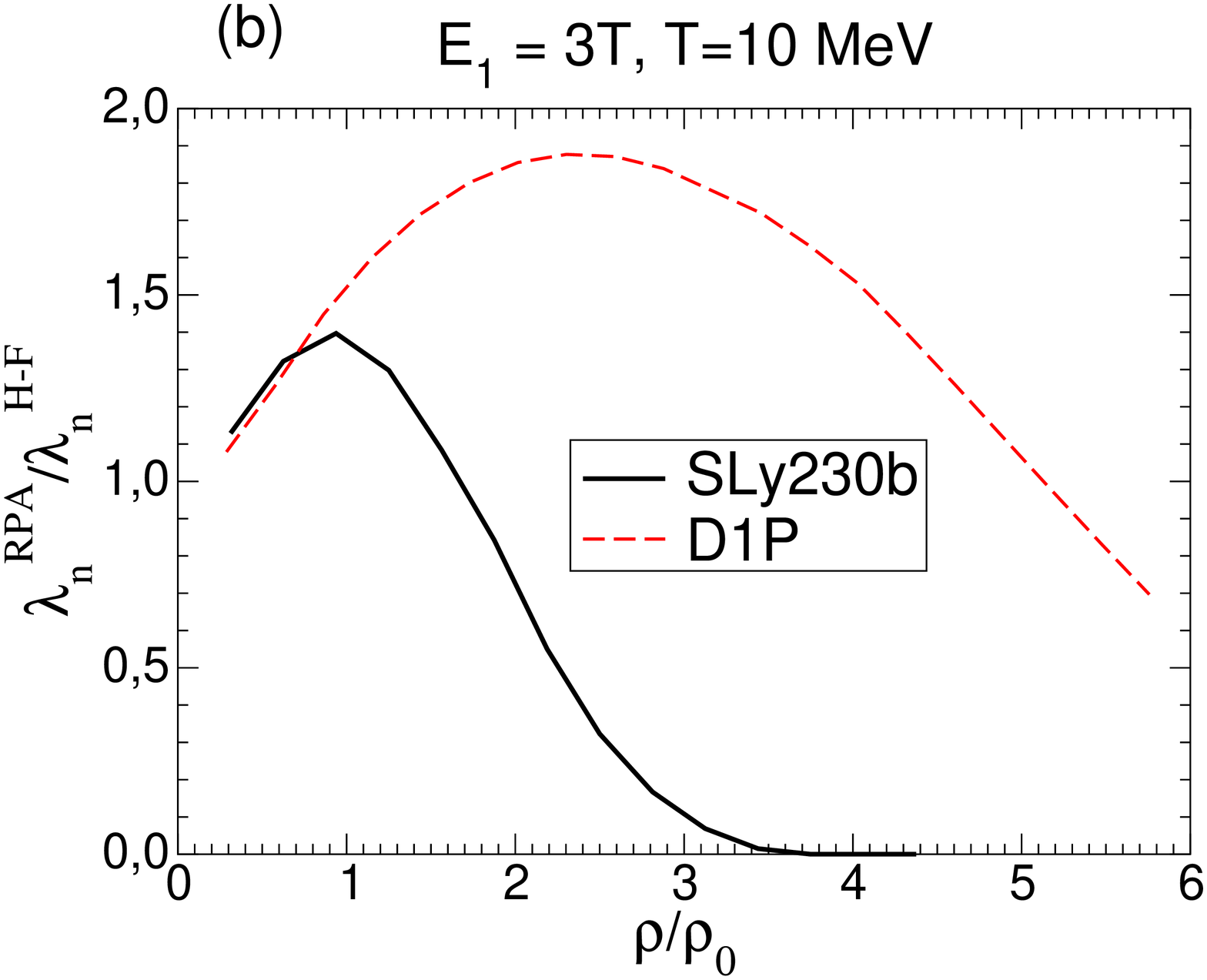}
\caption{Neutrino mean free path in pure neutron matter versus the
density. Several treatments of the nuclear correlations are shown:
the mean field approximation (a) and the 
RPA (b).}
\label{fig1}
\end{figure}
Fig. 1 displays our results for pure neutron matter. 
In Fig. \ref{fig1}(a), the neutrino mean free path is plotted versus 
the density for temperatures of 5 MeV and 20 MeV. In the case of a free 
gas with no nuclear interaction (solid line) the mean free path decreases, 
as expected. Mean field results are also plotted for the Skyrme 
SLy230a \cite{cha97} and Gogny D1P \cite{far99} interactions (dashed and 
dotted lines, respectively). The resulting wide range of mean free paths at
high density is directly related to the behaviour of the effective 
mass \cite{mar01b}. The effects of the long range RPA correlations are 
represented in Fig. \ref{fig1}(b). The ratio between the RPA 
($\lambda_n^{\rm RPA}$) and mean field ($\lambda_n^{\rm HF}$)
mean free paths is plotted as a function of the density. 
Around the saturation density, both interactions predict an increase
in the mean free path. However, as 
the density increases it turns out that spin 
fluctuations become dominant, reflected in the presence of a spin zero sound
mode which eventually produces a strong reduction of the mean free path. 

\smallskip
\begin{figure}[ht]
\centering
\includegraphics[angle=-90,scale=0.8]{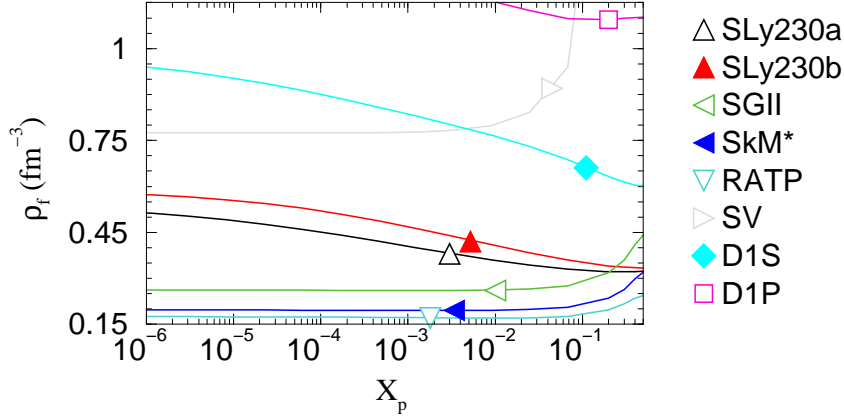}
\caption{Ferromagnetic phase diagram for Skyrme and Gogny effective 
interactions.}
\label{fig3}
\end{figure}

For Skyrme type interactions the mean free path vanishes at some
density beyond $\rho_0$, as a consequence of the appearence of spin
instabilities. Indeed, at high densities it is energetically favorable 
for nucleons to completely polarize their spins, and a ferromagnetic
transition is always predicted by these interactions. 
In Fig. \ref{fig3} we plot the 
critical density as 
a function of the proton fraction using several effective interactions. 
It can be seen that all Skyrme interactions predict 
ferromagnetism for densities below $\simeq 3.5 \rho_0$. 
In contrast, Gogny type interactions either predict this transition to
occur at much higher densities or not occur at all. 
If a ferromagnetic transition occurs at high densities, we may expect huge
modifications of the neutron star properties. For instance, we may observe
neutron stars  
with a ferromagnetic core and a normal neutron liquid around,
with a very huge magnetic field on its surface \cite{hae96}.
In this respect, several microscopic calculations based on realistic bare 
interaction have recently been performed in different frameworks as 
diffusion Monte Carlo \cite{fan01} or Brueckner-Hartree-Fock
\cite{vid02,lom02}.  
None of these microscopic calculations confirms the presence of a
ferromagnetic instability. This is a strong argument which casts some doubt
about the reliability of Skyrme effective interactions beyond saturation 
density.

\smallskip
We have performed a systematic study of all possible instabilities in 
symmetric nuclear matter and pure neutron matter predicted by effective 
Skyrme interactions \cite{mar02}. The stability conditions are those 
related to the inequalities that Landau parameters must satisfy. 
Of course, we impose the interactions to be consistent with currently accepted
values for such properties as binding energy, incompressibility, asymmetry 
energy and surface energy of symmetric nuclear matter at the saturation point.
\begin{figure}[ht]
\centering
\includegraphics[scale=0.5]{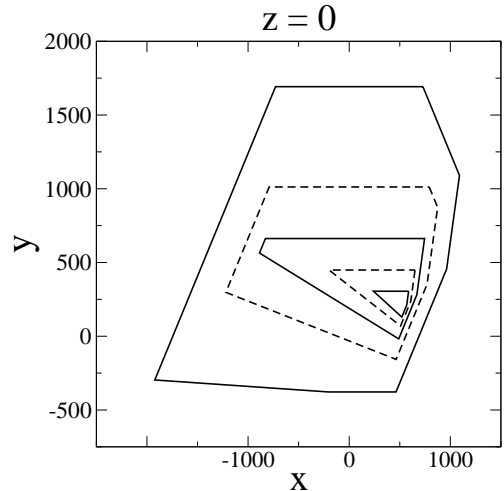}
\caption{Stability domains in the $(x,y)$ plane for $z=$0 and for several
densities: the largest area (external solid line) is for $\rho_0$, the next
one (dashed line) is for 1.5$\rho_0$, and so on in steps of 0.5$\rho_0$.}
\label{fig4}
\end{figure}
Starting from a general Skyrme interaction with ten parameters, it is possible
to show that seven parameters or combinations of parameters can be determined
by some of the above mentioned empirical inputs. The problem is thus reduced
to the study of instabilities in terms of three parameters, which we choose
as $t_1 x_1 \equiv x$, $t_2 x_2 \equiv y$, and $t_3 x_3 \equiv z$. 
For a given value of the density, the Landau inequalities define in the
$(x,y,z)$-space a volume or stability domain. Outside this domain, one
or more of the Landau inequalities are violated. As an example,
in Fig. \ref{fig4} are plotted
these domains in the $(x,y)$ plane for $z=$0 and 
several values of the density.
No instabilities appear for densities below the value associated to each
contour. The volume of the stability domain decreases when the density
increases, and there is a density where no solution exists satisfying 
the stability conditions. This figure shows that for any Skyrme interaction, 
there is a critical density beyond which nuclear matter is necessarily 
unstable.This result is robust because it is independent of the 
value of 
$z$ and of slight modifications of the
experimental constraints around their accepted values. The value of this 
critical
density does not exceed 3.5-4 times $\rho_0$ for a resonable choice of the
empirical inputs. The stability domains are well identified and it would 
be worthwhile 
to look inside them for Skyrme parametrizations that can also
describe accurately finite nuclei.

\smallskip 
As a conclusion, this discussion in an illustration that an accurate prediction
of the neutrino mean free path must also include an accurate understanding of
some nuclear properties like the k-effective mass and the onset of
instabilities.

\end{document}